\documentclass[]{aa} %
\usepackage{graphicx}
\usepackage{txfonts}
\usepackage{verbatim}
\usepackage{bm}

\newcommand{\figref}[1]{Fig.~\ref{#1}}
\newcommand{\eqnref}[1]{Eq.~(\ref{#1})}

\begin{document}
%

%Not sure about this title.  It's pretty good except a title really shouldn't have acronyms
   \title{The new NIKA: A dual-band millimeter-wave kinetic inductance camera for the IRAM 30-meter telescope}

%   \subtitle{I. Overviewing the $\kappa$-mechanism}

   \author{
   A. Monfardini\inst{1}\fnmsep\thanks{e-mail: monfardini@grenoble.cnrs.fr}, A. Benoit\inst{1}, A. Bideaud\inst{1},  L. J. Swenson\inst{1},
   M. Roesch\inst{2}, F. X. D\'esert\inst{3}, S. Doyle\inst{4}, A. Endo\inst{5}, A. Cruciani\inst{1,6}, P. Ade\inst{4}, A. M. Baryshev\inst{7}, J. J. A. Baselmans\inst{8}, O. Bourrion\inst{9}, M. Calvo\inst{6}, P. Camus\inst{1}, L. Ferrari\inst{7}, C. Giordano\inst{10}, C. Hoffmann\inst{1}, S. Leclercq\inst{3}, J. Macias-Perez\inst{9},
   P. Mauskopf\inst{4}, K. F. Schuster\inst{3}, C. Tucker\inst{4}, C. Vescovi\inst{9}, S. J. C. Yates\inst{8}
          }
         \institute{Institut N\'eel, CNRS \& Universit\'{e} Joseph Fourier (UJF), BP 166, 38042 Grenoble, France
         \and
         Institut de RadioAstronomie Millim\'etrique, 300 rue de la Piscine, 38406 Saint Martin d'H\`eres, France
         \and
         Laboratoire d'Astrophysique, Observatoire de Grenoble, BP 53, 38041 Grenoble, France
         \and
         Cardiff School of Physics and Astronomy, Cardiff University, CF24 3AA, United Kingdom
         \and
         Kavli Institute of NanoScience, Delft University of Technology, Lorentzweg 1, 2628 CJ Delft, The Netherlands
         \and
         Dipartimento di Fisica, Universit\'a di Roma La Sapienza, p.le A. Moro 2, 00185 Roma, Italy
         \and
         SRON, Netherlands Inst. for Space Research \& University of Groningen, Postbus 800, 9700 AV Groningen, Holland
         \and
         SRON, Netherlands Institute for Space Research, Sorbonnelaan 2, 3584 CA Utrecht, Holland
         \and
         Laboratoire de Physique Subatomique et de Cosmologie, UJF, CNRS/IN2P3, INPG, 38026 Grenoble, France
         \and
         Fondazione Bruno Kessler, Via Sommarive 18, I-38123 Povo (TN), Italy
             }

    \authorrunning{Monfardini et. al.}

   \date{Received December XX, XXXX; accepted XXX XX, XXXX}

% \abstract{}{}{}{}{}
% 5 {} token are mandatory

  \abstract
  % context heading (optional)
  % {} leave it empty if necessary
   {The N\'{e}el IRAM KIDs Array (NIKA) is a fully-integrated measurement system based on kinetic inductance detectors (KIDs) currently being developed for millimeter wave astronomy. In a first technical run, NIKA was successfully tested in 2009 at the Institute for Millimetric Radio Astronomy (IRAM) 30-meter telescope at Pico Veleta, Spain.  This prototype consisted of a 27-42 pixel camera imaging at 150 GHz.  Subsequently, an improved system has been developed and tested in October 2010 at the Pico Veleta telescope.  The instrument upgrades included dual-band optics allowing simultaneous imaging at 150 GHz and 220 GHz, faster sampling electronics enabling synchronous measurement of up to 112 pixels per measurement band, improved single-pixel sensitivity, and the fabrication of a sky simulator to replicate conditions present at the telescope. }
  % aims heading (mandatory)
   {The primary objectives of this campaign were:  to demonstrate imaging with the new dual-band optics, to evaluate the feasibility of hundred-pixel arrays of KIDs, to validate the sky simulator's ability to reproduce real observing conditions, to identify performance-limiting noise sources, and to image calibration and astronomically-relevant sources.}
  % methods heading (mandatory)
   {The imaging sensors consisted of two spatially-separated arrays of KIDs.  The first array, mounted on the 150 GHz branch, was composed of 144 lumped-element KIDs.  The second array, mounted on the 220 GHz branch, consisted of 256 antenna-coupled KIDs with a distributed resonator geometry. Each of the arrays was sensitive to a single polarization; the band splitting was achieved by using a grid polarizer. The optics and sensors were mounted in a custom dilution cryostat, with an operating temperature of $\sim$70 mK.  Electronic read-out was realized using frequency multiplexing and a transmission line geometry consisting of a coaxial cable connected in series with the sensor array and a low-noise 4 K amplifier.}
  % results heading (mandatory)
   {The new dual-band NIKA was successfully tested in October 2010, performing in-line with sky simulator predictions.  Initially the sources targeted during the 2009 run were re-imaged, verifying the improved system performance.  An optical NEP was then calculated to be around $2 \cdot 10^{-16}$ W$/$Hz$^{1/2}$.  This improvement in comparison with the 2009 run verifies that NIKA is approaching the target sensitivity for photon-noise limited ground-based detectors.  Taking advantage of the larger arrays and increased sensitivity, a number of scientifically-relevant faint and extended objects were then imaged including the Galactic Center SgrB2(FIR1), the radio galaxy Cygnus A and the NGC1068 Seyfert galaxy.  These targets were all observed simultaneously in the 150 GHz and 220 GHz atmospheric windows.}
  % conclusions heading (optional), leave it empty if necessary
   {}


   \maketitle
%
%________________________________________________________________

\section{Introduction}

The importance of millimeter and submillimeter astronomy is now well established. In particular, three main areas of millimeter continuum research have motivated the rapid development of new technologies:
\begin{enumerate}{}{}
\item The study of star forming regions in the Galaxy (\cite{Ward2007}). The pre-stellar phases in molecular clouds are hidden by cold dust (around 10 K) which can only be observed at submillimeter wavelengths.
\item The investigation of high redshift galaxies (\cite{Lagache2005}).  The redshift effect at submillimeter wavelengths counteracts the distance dimming (\cite{Blain2002}).
\item The measurement of cosmic microwave background (CMB) temperature anisotropies (either primordial or secondary).  At a temperature of 2.725 K, the CMB spectrum peaks at millimeter wavelengths. Of particular interest is the Sunyaev-Zel'dovich effect (\cite{Birkinshaw1999}), which distorts the CMB spectrum at millimeter and radio wavelengths and can be used to map the distribution of hot gas in clusters of galaxies.
\end{enumerate}
Throughout the previous decade, instruments utilizing hundreds of individual bolometers in focal plane arrays have dominated continuum submillimeter and millimeter astronomy (e.g. MAMBO2, BOLOCAM). Full-sampling arrays with up to thousands of pixels in a single array are now reaching maturity, offering increased mapping speed and decreased per-pixel manufacturing costs (e.g. Apex-SZ, SPT, SCUBA2, LABOCA).  Despite these considerable advances, further array scaling is strongly limited by the multiplexing factor of the readout electronics.

% Kids summary description
A promising alternative to traditional bolometers is the kinetic inductance detector (KID).  First demonstrated less than 10 years ago (\cite{Day2003}), a KID consists of a high-quality superconducting resonant circuit electromagnetically coupled to a transmission line.  Typically, a KID is designed to resonate from 1 to 10 GHz and exhibits a loaded quality factor exceeding $Q_L>10^5$.  Thus each KID occupies a bandwidth of order $\Delta f = f/Q_L \sim$10-100 kHz.  A single KID only loads the transmission line within $\Delta f$ around its resonant frequency.  The KID resonant frequency can easily be controlled geometrically during the circuit design.  It is therefore possible to couple a large number of KIDs to a single transmission line without interference as long as the inter-resonance frequency spacing exceeds $2\Delta f$.  The inter-resonance frequency spacing and the total bandwidth of the measurement electronics sets a limit on the total number of pixels which can be simultaneously read out on a single measurement line (\cite{Mazin2004}).  Current frequency-multiplexing measurement electronics are limited to at most a few hundred pixels per measurement cable.  This is expected to grow into the thousands of pixels per cable within the next decade, vastly improving the multiplexing factor in comparison with existing array technologies.

In order to absorb incident radiation, it is necessary to impedance match the KID to free space at the target wavelength.  Currently, there are two methods for meeting this criteria.  The first is to use a geometry known as the Lumped-Element KID (LEKID) which separates the resonator into an inductive meander section and a capacitor.  At the target wavelength, the inductive meander approximates a solid absorber.  By accounting for the resonator and substrate impedances and the cavity formed with the sample holder, it is possible to directly impedance match the LEKID to free space (\cite{doyle:156}).  Alternatively, a lens and antenna structure can be used to adapt the resonator to free space.  Due to the increased geometric complexity, it has proven more difficult to achieve satisfactory device performance with antenna structures than with direct absorbtion LEKIDs.  Despite this drawback, antenna-coupled KIDs remain a very active area of research due to their frequency selectivity (\cite{Schlaerth2008}).

For both LEKIDs and antenna-coupled KIDs, detection is achieved in the same manner once the incident radiation has been absorbed.  In a superconductor, the conduction electrons are condensed into charge carrying Cooper pairs.  Mediated by lattice vibrations, superconductivity results in an energy gap in the carrier density of states.  Incident photons with an energy exceeding the gap energy can break a Cooper pair, producing two quasiparticles and a concurrent change in the complex impedance.  The result is a shift in the KID resonance frequency which can be read out by the measurement electronics.

%NIKA summary and overview of article
We are currently developing a fully-integrated measurement system based on KIDs known as the N\'{e}el IRAM KIDs Array (NIKA).  The two primary goals of NIKA are to asses the viability of KIDs for terrestrial astronomy and to develop a filled-array, dual-band resident instrument for the Institute for Millimetric Radio Astronomy (IRAM) 30-meter telescope at Pico Veleta, Spain.  Based on a custom-designed dilution cryostat with a base temperature of $\sim$70 mK (\cite{benoit:702009}), a first generation single-band NIKA prototype was previously tested at the Pico Veleta telescope in October 2009 (\cite{Monfardini:29}).  This successful measurement was the first to directly compare the performance of LEKID and antenna-coupled KID designs.

Leveraging the experience gained from the first generation NIKA, an improved instrument has been designed and tested at the IRAM 30-meter telescope in October 2010.   This second generation system includes a large number of enhancements.  Dual-band optics, integrating a polarization-sensitive splitter and a new baffling structure, allow simultaneous imaging at 150 GHz and 220 GHz.  Resonator design modifications resulted in improved single-pixel sensitivity.  Faster digital-signal-processing electronics enable synchronous measurement of up to 112 pixels for each measurement band.  Fabrication of a sky simulator to replicate typical measurement conditions at the telescope facilitated improved array testing and quality control.  Along with a detailed discussion of these system upgrades, we present here the results of the October 2010 measurement campaign.  This includes a discussion of the limiting noise sources, an analysis of the system performance determined using calibration sources such as planets, and an estimate of the improved full-system sensitivity.  Finally we present astronomically-relevant observations of a number of faint and extended sources in both measurement bands which were previously unattainable with the less sensitive first generation NIKA.

%__________________________________________________________________

\section{The dual-band optics}

\begin{figure}[h]
  \centering
   \includegraphics[width=.95\linewidth]{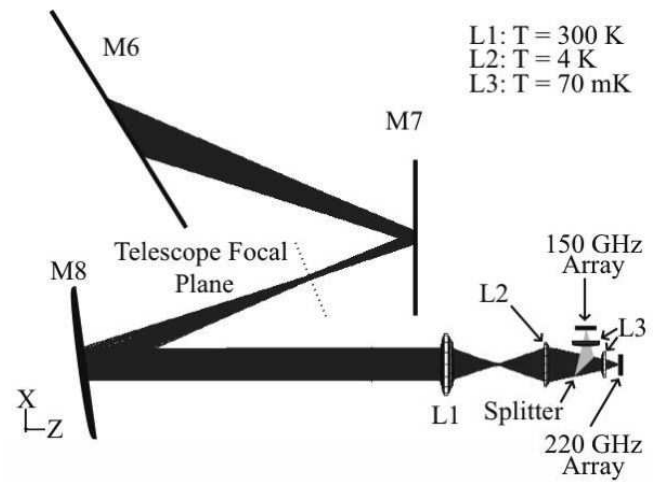}
      \caption{NIKA optical design. Two flat mirrors (M6,M7) orient the beam on the bi-conic mirror M8. The rays enter the cryostat through the window, coincident with lens L1. A grid polarizer located at 70 mK between lenses L2 and L3 splits the beam and redirects the radiation to two polarization-sensitive arrays, the first optimized for 150 GHz and the second for 220 GHz. The band-defining filters centered at 150 GHz and 225 GHz are mounted directly on the flat face of each L3 lens. }
         \label{fig1}
\end{figure}

%Large optics
The new dual-band NIKA is engineered to fit the receiver cabin of the IRAM telescope in Pico Veleta, Spain. The 30-meter primary mirror (M1) and the hyperbolic secondary (M2, D=2m) are installed directly on a large alt-azimuth mounting.  The incident beam is directed into the receiver cabin through a hole in M1 using a standard Cassegrain configuration.  A rotating tertiary (M3) provides a fixed focal plane (Nasmyth focus). The optical axis, in order to conform to the dimensions of the cabin, is deviated by two flat mirrors (M4 and M5). M4 can rotate between two fixed positions, selecting either heterodyne or continuum instruments.

%Cryo optics
NIKA re-images the large telescope focal plane onto the small sensitive area covered by the KIDs. The demagnification factor is around 6.6, achieving a well-adapted scale of 5 arcseconds/mm on the detector plane. This is accomplished using two flat mirrors (M6, M7), one bi-conical mirror (M8) and three high-density polyethylene (HDPE) corrugated lenses (L1, L2, L3). The lens corrugation consists of machined concentric grooves, providing a soft transition between vacuum (n=1) and HDPE (n=1.56) to reduce reflective losses. The size of the grooves is 0.4 mm $\times$ 0.4 mm in depth and width, while the width of the ridges is 0.4 mm.  L1 is located at room temperature and coincides with the cryostat vacuum isolation window. L2 is mounted on the screen at 4 K, while the final L3 lenses (one per array) are installed at the coldest stage just in front of the arrays ($\sim$70 mK).  A simple grid polarizer with a grid pitch of 4 ${\mu}$m was inserted at 45 degrees with respect to the main optical axis before the final L3 lenses. Since all KID designs currently employed in NIKA are sensitive to a single polarization, this is an efficient way to realize beam splitting for the two frequency bands.  A schematic of the optical design is presented in \figref{fig1}.

%Array details
For both arrays, the lenses are telecentric.  That is, each point of the detector plane is illuminated with the same aperture with the chief ray perpendicular to the surface.  This strongly reduces potential systematic errors related to overlap of the incident beam with the intrinsic pixel lobe. The detector plane is over-sampled with respect to the diffraction spot size: the full width at half maximum is about 3 mm for the 150 GHz band and 2 mm for the 220 GHz band. The pixel pitches are 2.25 mm and 1.6 mm, for final aperture ratios of 0.75$\cdot$F$\lambda$ and 0.80$\cdot$F$\lambda$ respectively. The useful, projected field-of-view is approximately $1.6 \times 1.6$ arcminutes for the 150 GHz array and $1 \times 1$ arcminutes for the 220 GHz array.

%should that have been arcminute^2 in the last paragraph for fov?

\begin{figure}[h]
   \centering
    \includegraphics[width=.95\linewidth]{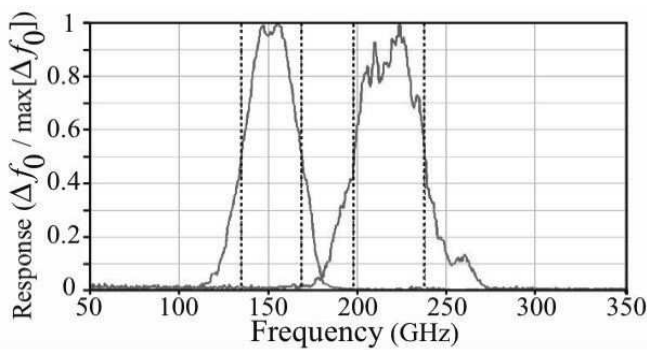}
      \caption{Normalized spectral response of the NIKA detectors. The 3 dB measured bandwidths are 135-169 GHz and 198-238 GHz.}
         \label{fig2}
\end{figure}

%Filtering and baffling
%The Back Focal Length (BFL) is 28~mm.
A series of filters and baffling are used to reject unwanted radiation.  To reduce off-axis radiation, a multi-stage, black-coated baffle is installed at 4 K between L1 and L2.  This is further augmented with a cold pupil at 70 mK.  Three infrared-blocking filters are mounted on the first two radiative screens held at a temperature of $\sim$150 K and $\sim$70 K by cold helium vapors. Three additional low-pass filters are installed on the 4 K and 1 K screens.  The radiation entering the 70 mK stage is thus restricted to frequencies $\nu<$300 GHz. The final band definition is achieved using a series combination of a high-pass and a low-pass filter behind the final lens L3, just in front of the detector array.  The spectral response of NIKA was characterized using a Martin-Puplett interferometer (MPI) (\cite{Durand:2007}). The results, displayed in \figref{fig2}, exhibit good agreement with the atmospheric transparency windows and coincide with the expected response based on the individual filter cutoff frequencies.  Taking into consideration the tabulated HDPE transmission and the individual filter specifications, we estimate a total optical transmission coefficient of $\approx$0.4 for the NIKA optics.

%______________________________________________________________

\section{The detectors and readout electronics}

\begin{figure}[h]
    \centering
    \includegraphics[width=.95\linewidth]{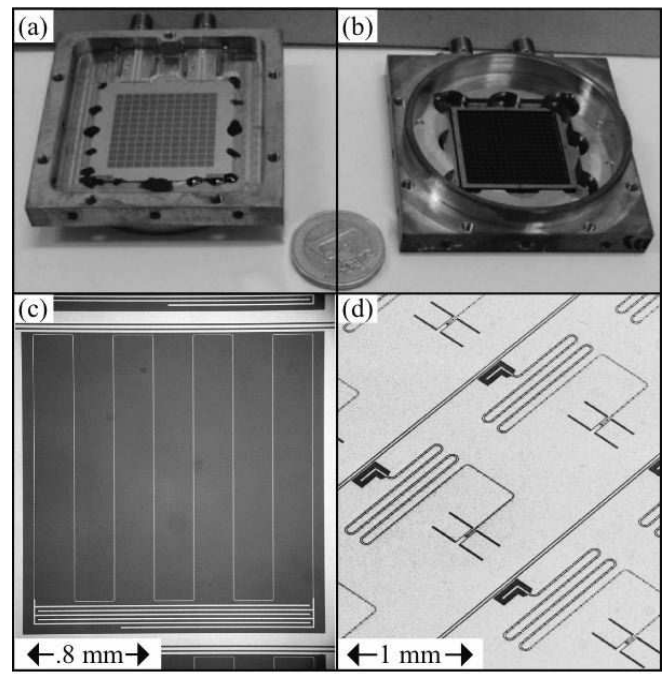}
    \caption{Mounted arrays and pixel micrographs.  (a) 144 pixels LEKIDs array.  (b) 256 pixels antenna-coupled array.  KID micrographs: (c) LEKID and (d) Antenna-coupled KID.  For both micrographs, light regions indicate where metal is present and dark areas where the substrate dielectric is exposed.}
     \label{fig3}
\end{figure}

%Lekid fabrication
The second generation NIKA implements a 144 pixel array of LEKIDs for 150 GHz detection and a 256 pixel array of antenna-coupled KIDs for 220 GHz sensing. An image of the mounted arrays is shown, along with micrographs of the individual pixels, in \figref{fig3}.  For the LEKIDs array, fabrication commenced with an argon plasma surface treatment of a 300 ${\mu}$m thick high-resistivity silicon wafer ($>$5 k$\Omega$/cm).  Next, a 20 nm aluminum film was deposited via DC sputtering.  The final array structure was subsequently defined with UV lithography followed by wet etching.

% Antenna-coupled fabrication
For the antenna-coupled KIDs, a hybrid material structure necessitated a more complex fabrication strategy.  The basic KID consists of a quarter-wave coplanar waveguide (CPW) resonator.  One end of the resonator is coupled to the measurement transmission line while the other end is grounded.  A twin-slot antenna focuses incident radiation on the grounded end.  Half the length of the resonator center strip, on the grounded end, is made of aluminum; the rest of the center strip on the coupler side, and the entire ground plane, is made of NbTiN.  Fabrication consisted of a 300 nm thick sputter deposition of NbTiN (\cite{barends2010}) on a HF-passivated high-resistivity silicon wafer ($>$10 k$\Omega$/cm).  A 100 nm aluminum film was then sputter deposited through a lift-off mask into the plasma-etched slots in the ground plane to complete the resonator center strip.  Aluminum air-bridges were then used to connect the ground planes of the CPW in order to suppress microwave mode conversion, reduce the cross talk and improve the beam pattern.  The radiation is concentrated and focused on the antennas by silicon elliptical lens segments of 1.6 mm diameter, arranged in a rectangular grid of 1.6 mm spacing.  This lens array is glued on the back side of the KID sample. The misalignment between the lens and the twin slot antenna is less than 10 $\mu$m, ensuring good coupling efficiency.

% Roach electronics
The general principles of frequency-multiplexed KID readout have been described in detail elsewhere (\cite{swenson:84}, \cite{yates:042504}, \cite{Monfardini:29}). The current digital electronics used for the NIKA readout were developed in the context of an international collaboration known as the Open Source Readout (OSR) (\cite{duan:7741}).  This system was based on a digital platform known as ROACH, itself having been developed in the context of another collaboration known as Center for Astronomy Signal Processing and Electronics Research (CASPER) (\cite{Parsons2006}).  The ROACH hardware provides powerful signal processing capabilities by integrating a field-programmable gate array, an on-board power pc, and a variety of high-speed communication interfaces.  Building on this, the OSR developed new high-speed, dual-input analog-to-digital and dual-output digital-to-analog interface cards.  For NIKA, both of these cards were clocked by the same rubidium-referenced external clock generator.  Operating at 466 megasamples per second, the resulting useful IF measurement bandwidth of the NIKA readout was 233 MHz.  In order to drive the individual pixels and subsequently read out their state, the NIKA collaboration developed customized software to use with the OSR hardware.  Similar to standard lock-in techniques, the implemented algorithm allowed 112 separate measurement tones to be generated and simultaneously monitored within the IF measurement bandwidth.

\begin{figure}[h]
    \centering
    \includegraphics[width=0.95\linewidth]{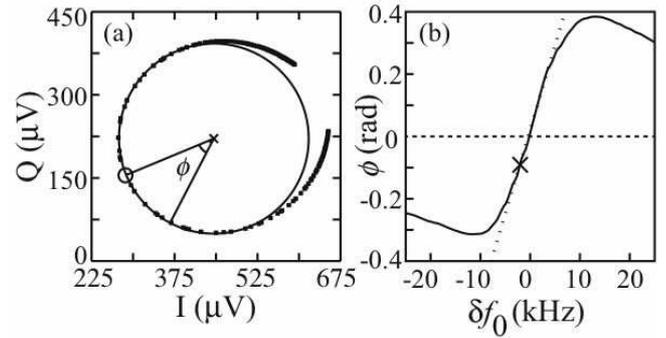}
    \caption{KID calibration.  (a) I-Q frequency scan across a resonance. Complex phase is calculated by fitting each resonance to a circle (shown); the phase angle is then determined with respect to the center of the resonance circle as indicated by the x. The minimum transmission is marked with a small circle.  (b) Frequency shift versus complex phase. The x on the curve indicates the maximum frequency shift when Mars transits the pixel.}
     \label{fig4}
\end{figure}

% Measurement band and excitation power
The NIKA readout uses a standard up-down converter configuration based on two IQ mixers per board to transpose the generated IF frequency comb to the resonator operating frequencies.  One ROACH board and a set of IQ mixers was used for each array.  Currently, the LEKID array operates in the frequency range 1.27-1.45 GHz.  Within this bandwidth, 104 pixels and 8 off-resonance blind tones could be used for measurement.  The antenna-coupled KID array has a central pixel core operating at 5-5.2 GHz.  Due to the larger inter-resonator frequency spacing of this array, only 72 core pixels could be simultaneously measured out of the total 256 pixels in the array.

The response of every pixel in an array is measured simultaneously and broadcast via UDP packets by the ROACH electronics to the control computers at a rate of 22 Hz.  The individual pixel responses are composed of a pair of in-phase ($I$) and quadrature ($Q$) values which result from the final stage of digital mixing and low-pass filtering.  Theses values can be translated into the traditional transmission phase and amplitude using the identities $\theta$ = $\arctan(Q/I)$ and amplitude$^2$ = $I^2$ + $Q^2$.  An alternative approach is to plot these values in the complex plane.  An example of a standard frequency sweep around a resonance is provided in \figref{fig4}(a).  Small changes in illumination result primarily in motion around this curve and thus it is convenient to define a new angle $\phi$ about the center of curvature $(I_c, Q_c)$:
\begin{equation}
\label{1}
\phi =\arctan \left ( \frac{Q-Q_{c}}{I-I_{c}} \right )-\phi_{0}
\end{equation}
where $\phi_0$ rotates the plane such that the curve intersects the x axis at the resonance frequency $f_0$.

\begin{figure}[h]
    \centering
    \includegraphics[width=0.95\linewidth]{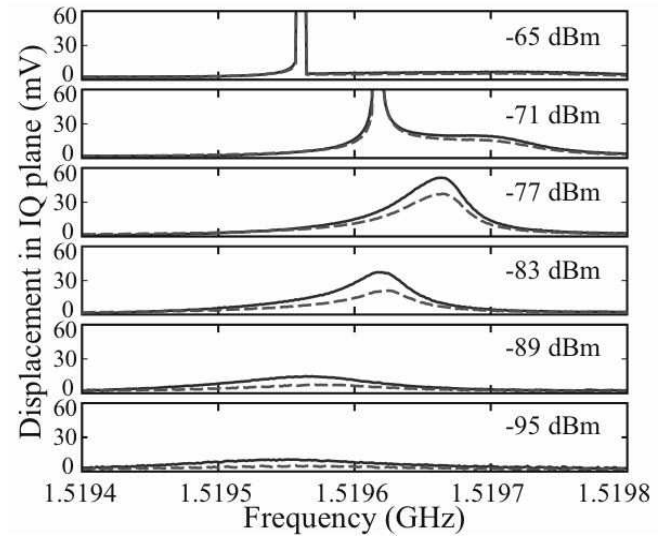}
    \caption{Readout power and frequency optimization.  The velocity magnitude $[(\Delta I / \Delta f)^2 + (\Delta Q / \Delta f)^2]^{1/2}$ (solid, blue) is plotted along with the optical response $[(\Delta I / \Delta T)^2 + (\Delta Q / \Delta T)^2]^{1/2}$ (dashed, red) for each point around a resonance.  The velocity was measured while making linear frequency steps $\Delta f =$ 6 kHz around the resonance.  The optical response was measured by sweeping the resonance in frequency with a cold focal plane ($T \sim 70$) K and then at increased temperature $\Delta T = $7 K.}
     \label{fig5}
\end{figure}

We have observed that the KID response to radiation depends critically on the driving power and frequency of the readout electronics.  While still not fully understood, it is reasonable to conjecture that this dependence could be caused by: non-linear thin film effects, quasiparticle generation or suppression in the superconductor, or a field-dependent dielectric behavior.  A plot of the velocity around the resonance curve for a linear frequency sweep and the measured response to a change in the focal plane temperature is shown in \figref{fig5} for increasing readout powers.  For low readout power ($<$ 90 dBm), the responsivity of the KID is weak.  As the power is increased, the KID responsivity and the velocity are seen to increase and to shift to higher frequencies.  Beyond a critical readout power, a discontinuity in the resonance emerges and the frequency shift changes directions and moves back toward lower frequencies.  While the greatest responsivity is obtained in this region, the region near the discontinuity is unstable and not currently well understood.  For the current measurement a readout power of $\sim -77$ dBm, just below the appearance of the discontinuity, was used.  For the frequency selection, it is also important to note that while the maxima in responsivity and velocity coincide, these generally do not occur near an extrema of the traditional $S_{21}$ amplitude or phase or their corresponding derivatives.  The exact relationship between the responsivity and the shape of the $S_{21}$ curve is observed to depend on a number of implementation details, including pixel cross-talk and impedance loading of the transmission line.  For this reason, the NIKA software now implements a calculation of the velocity around the resonance during calibration.

For low intensity radiation, the density of photo-generated quasiparticles in a KID is proportional to the incident photon flux.  To first order, this results in a linear shift in the kinetic inductance and, for thin films, a subsequent linear shift in the resonance frequency (\cite{SwensonAPL96}).  That is:
\begin{equation}
\label{2}
\delta f_{0}=-Cf_{0}^{3}\delta L_{K}\propto -\frac{f_{0}^{3}}{n_{s}^{2}} \delta P_{i}
\end{equation}
%Actually, what is n(s) here?
where $C$ is a constant, $n_{s}$ is the Cooper pair density, and $\delta P_{i}$ is the incident power.  A plot of the phase $\phi$ versus the frequency shift from resonance $\delta f_0$ for a typical KID is shown in \figref{fig4}(b) along with the maximum frequency shift during a transit of Mars.  From this plot it is clear that $\partial\phi/\partial f$ is approximately linear for relevant astronomical signals. Thus $\phi \propto \delta P_{i}$ with the constant of proportionality being determined during a calibration scan.  One caveat is that under large background changes, which can result from weather changes or telescope repointing, $n_s$ changes substantially and causes a significant shift in $f_0$.  For this reason, an automatic procedure was implemented to properly determine $(I_c, Q_c)$ and $\phi_0$ before every scan taken on the telescope to assure that the measurement would remain in the dynamic range of the KIDs.  This recalibration is currently achieved in under 60 seconds.

%______________________________________________________________

\section{The sky simulator}

\begin{figure}[h]
    \centering
    \includegraphics[width=.95\linewidth]{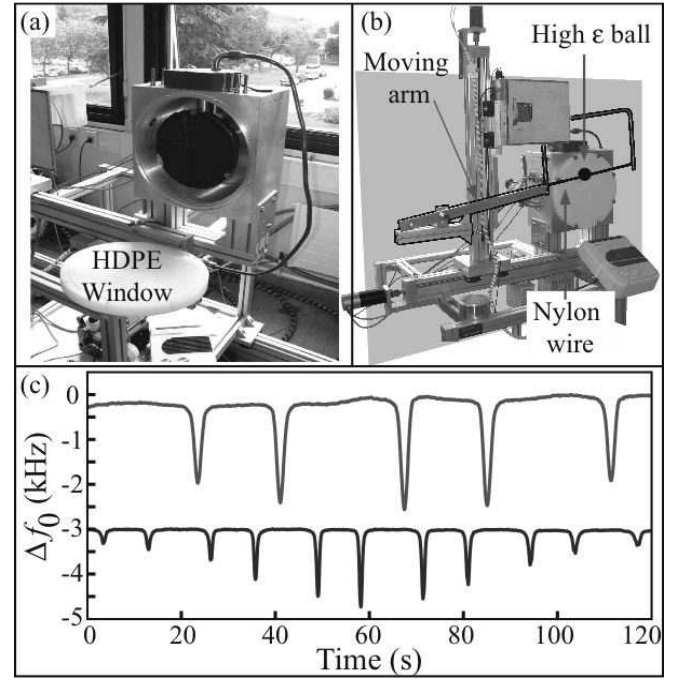}
    \caption{Sky simulator.  (a) HDPE cryostat window removed. The cold disk simulating the sky background is shown; its diameter is 24 cm.  (b) Fully assembled system.  The translation arm, nylon wire and high emissivity ball are outlined for visibility.  The ball moves at a controlled rate in front of the HDPE window, faking a real source on the sky. In particular, on-the-fly telescope scans are easily simulated (\cite{Bideaud:2010}). (c) Sky simulator validation.  A real on-the-fly scan of Uranus taken at the 30-meter telescope is shown (red, upper) along with a laboratory on-the-fly scan using the sky simulator (blue, lower). The telescope angular scan speed was 10 arcseconds/s. The linear scan speed of the sky simulator was 30 mm/s, corresponding to a telescope angular speed of 20 arcseconds/s. For both traces, the transit of the source was observed with a single pixel.}
     \label{fig6}
\end{figure}

%General description
In order to replicate real observing conditions and to properly estimate the amount of stray-light on the detectors, we have built a testing tool to complement MP interferometer measurements and classical chopper tests alternating between hot and cold sources.  Simply called the sky simulator, the basic idea consists in cooling down a large, black disk with the same dimensions as the telescope focal plane.  This cold disk simulates the background temperature in ordinary ground-based observing conditions.  On a telescope the main contributions to the background are the atmospheric residual opacity, which is weather dependent, and the emissivities of the mirrors.  The typical background temperature of the IRAM Pico Veleta telescope is in the range 30-100 K.  For the sky simulator, the cooling of the background disk is achieved by using a single stage pulse-tube refrigerator. A large window in the sky simulator cryostat is fabricated in 4 cm thick HDPE which is sufficiently transparent at millimeter wavelengths. The minimum background temperature that can be achieved is 50 K, limited by the radiation absorbed by the large, black, cold disk.  An image of the sky simulator can be seen in \figref{fig6}.

% Ball and simulation
To simulate an astronomical source, a high-emissivity ball with a diameter of 5-10 mm was placed in front of the sky simulator window at room temperature.  The ball is mounted on a motorized XY stage enabling movement with respect to the fixed disk.  The angular speed on the sky of a typical on-the-fly scan at the IRAM telescope is 10 arcseconds/s.  Accounting for the $\sim$300 m effective focal length of M1 and M2, this corresponds to a scan speed of 15 mm/s on the telescope focal plane.  This velocity is well within the capabilities of the XY translator.  A comparison of two single-pixel transits, the first taken with the telescope and the second with the sky simulator, is shown in \figref{fig6}(c).

% Sky simulator noise estimation
The sky simulator temperature can be adjusted continuously between 50 K and 300 K, allowing an accurate estimation of the detector response and a direct determination of the noise equivalent temperature (NET).  To perform this measurement, the spectral noise density $S_n(f)$, expressed in Hz/Hz$^{1/2}$, is calculated at a fixed background temperature.  The background temperature is then increased by ${\delta}T$ and the shift in the KID resonance frequency is measured.  This yields the KID frequency response $R$ expressed in Hz/K.  The NET is then given by:
\begin{equation}
\label{3}
NET(f) = \frac{S_{f}(f)}{R}
\end{equation}

% Sky simulator NEP/NET
Laboratory tests performed on the LEKID array indicated an average NET per pixel of 4 mK/Hz$^{1/2}$ at a standard representative frequency of 1 Hz. Accounting for the optical chain transmission and including a 50\% reduction due to the polarizer, this corresponds to an optical noise equivalent power (NEP) of approximately $2 \cdot 10^{-16}$ W$/$Hz$^{1/2}$. The best single-pixel LEKIDs have an optical NEP, under realistic 4 to 8 pW loading per pixel, around $6 \cdot 10^{-17}$ W$/$Hz$^{1/2}$.  These results indicate that no conceptual limitations exist for using KIDs in the next generation large arrays of ground-based mm-wave instruments.

% Stray light/baffling test
The sky simulator has also been used to estimate the amount of undesired stray light incident on the detector arrays.  This measurement required two steps.  First the pixel frequency shifts were recorded while the sky simulator was moved in the optical axis direction from its customary position at the telescope focal plane up to the cryostat window.  Next, the sky simulator was returned to the telescope focal plane and the background temperature was then increased until an equivalent frequency shift was affected.  This resulted in an estimated stray-light temperature of $\sim 35$ K which corresponds to $\sim 4$ pW of parasitic power per pixel at 150 GHz. The unwanted radiation has thus been reduced by more than a factor of two compared with the first generation NIKA and is now comparable to the best sky conditions at Pico Veleta.

%______________________________________________________________

\section{Results}

The dual-band NIKA run took place in October 2010. The instrument was installed in the receiver cabin of the IRAM 30-meter telescope at Pico Veleta, Spain, and operated remotely from the control room.  The cool-down of the instrument was also performed remotely, taking approximately 18 hours to reach the operating temperature of 70 mK.

\begin{figure}[h]
    \centering
    \includegraphics[width=.95\linewidth]{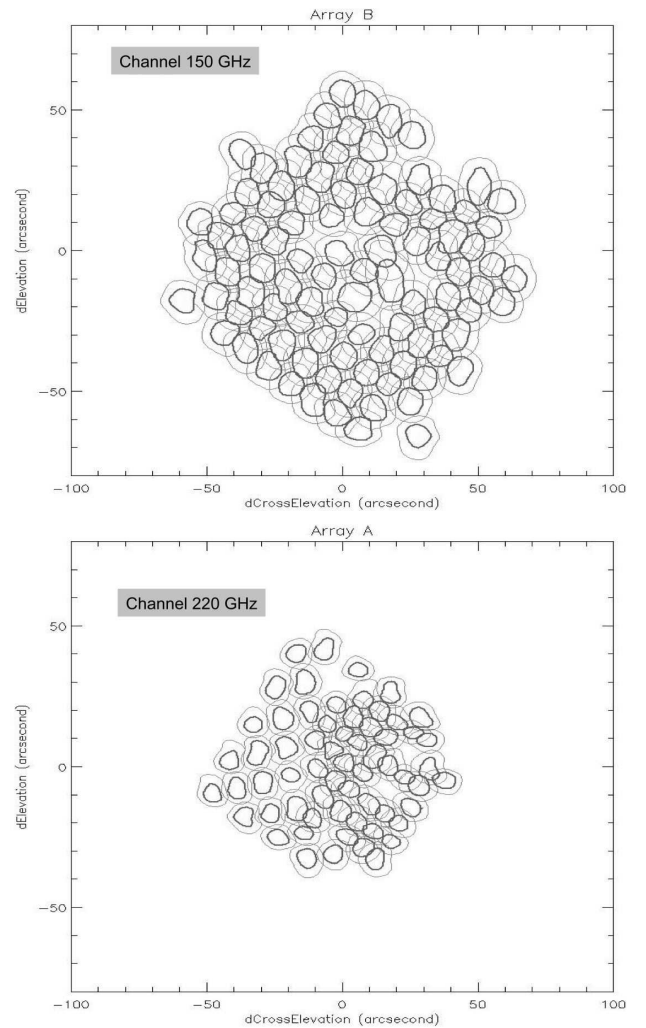}
    \caption{$80\%$ (Blue) and $50\%$ (Grey) contours of the beams as measured on Mars on the 2010-Oct-20. Top: 2~mm array (98 valid KID), Bottom: 1.4~mm array (62 valid KID). }
     \label{Fig7}
\end{figure}

%START XAVIER CONTRIBUTION
Astronomical data from the two arrays are reduced off-line with dedicated software. The raw data ($I$,$Q$) are converted to complex phase angle using the closest previous KID calibration. Then a conversion to an equivalent frequency shift is done with the same calibration using the derivative of the frequency with the complex phase at the zero phase, as described in \figref{fig4}. Data are thus internally converted to frequencies which are assumed to be linear with the absorbed photon counts, as in equation \eqnref{2}. After opacity correction, and using Mars as the primary calibrator, we obtain that the overall median gain is of 14 mJy/beam/Hz and 9 mJy/beam/Hz for the 1.4 and 2~mm (220 GHz and 150 GHz) channels, with a $30\%$ dispersion. The focal plane geometry of each array is measured by using scanning maps of planets (see \figref{Fig7}).

The fitted focal plane geometry is found by matching the pixel position in the
array as measured on the wafer to the measured position on planets, by
optimizing a simple set of parameters: a center, a tilt angle and a scaling
expressed in arcseconds/mm. Most detectors are within less than 2 arcseconds
of their expected position. The beam width is also found from planet
measurements. Typically the FWHM is 12.4 and 16.7 arcseconds for the two arrays (1.4 and 2~mm respectively, see figure 8)
with a dispersion of 1 arcsecond. This is close to the diffraction limit for
the 2~mm array. Pixelisation effects have not been removed from this
estimate. This might explain why the 1.4~mm beam is larger by 20\% than the
diffraction size.

\begin{figure}[h]
    \centering
    \includegraphics[width=.95\linewidth]{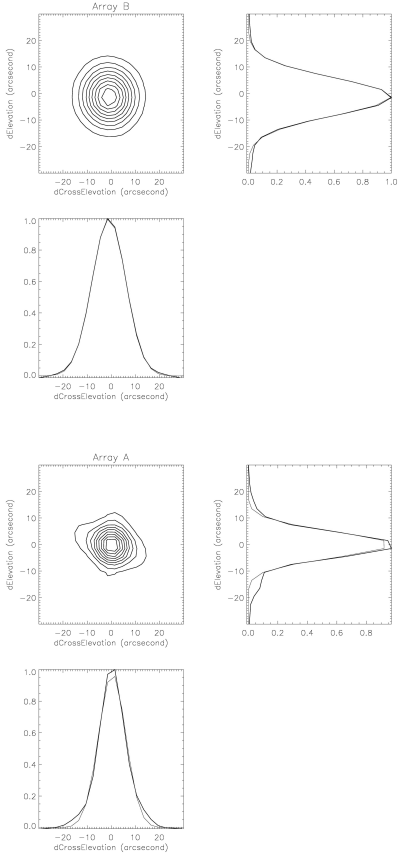}
    \caption{Caption: Beam contours obtained on Mars with levels at 10 to 90 percent.
Azimuth and elevation slices along the center are shown on the sides
along with the slice of a 2D Gaussian fit. The FWHM is 16.7 (resp. 12.4)
arcseconds at 2~mm (resp. 1.4~mm).}
     \label{Fig8}
\end{figure}

We were able to observe many types of objects. We here show only examples of
the results that are currently analyzed. Usually, the observing scan mode is a simple
on-the-fly mapping, composed of constant-elevation sub-scans, with the
telescope sweeping in azimuth back and forth. The detector data are processed
in the time domain (calibration and filtering) before being projected on a
common map, using the focal plane geometry as described above. The noise is
evaluated at the detector level by histogram Gaussian fitting outside the
sources. The noise is then propagated to the map level. The final map is
obtained by coadding the individual detector maps with an inverse square noise
weighting scheme. From known sources and reproducibility from scan to scan and
from detector to detector, the photometric accuracy is estimated to be 30\%. New methods are being investigated to improve this
photometric offline processing accuracy.

We present SrgB2(FIR1) as an example of extended source, Cygnus A as an
example of a multiple source, and NGC1068 as an example of weak source with a
new 2~mm flux measurement.

\begin{figure}[h]
    \centering
    \includegraphics[width=.95\linewidth]{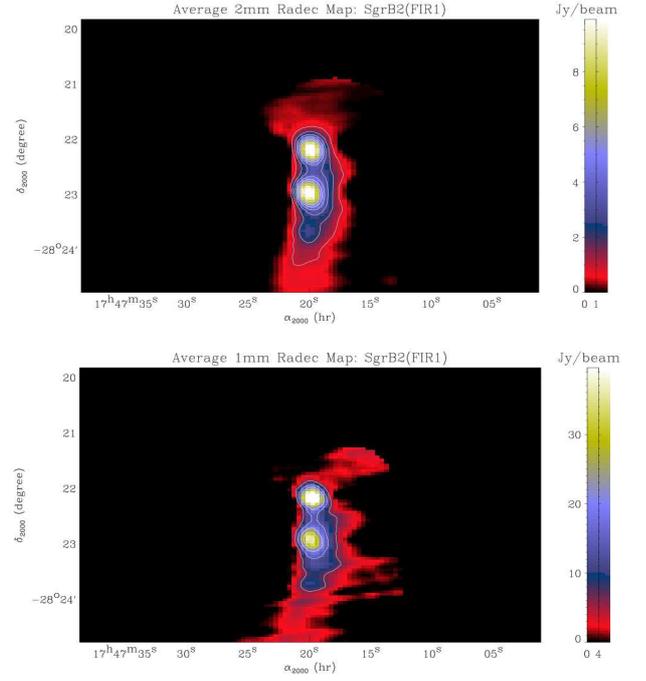}
    \caption{SgrB2(FIR1) maps at both frequencies. The Galactic Center map is
      complex and shows at least three common sources and one North-South extended
      component that does not coincide in the two maps.}
     \label{Fig9}
\end{figure}

\figref{Fig9} shows a dual map of the Galactic Center obtained in 900
seconds of integration time. This complex region reveals at least 3 compact
sources. The very center has a flux of $76\pm3,\mathrm{Jy}$ and $17.7
\pm{0.7},\mathrm{Jy}$ at 1.4 and 2~mm, respectively.

\begin{figure}[h]
    \centering
    \includegraphics[width=.95\linewidth]{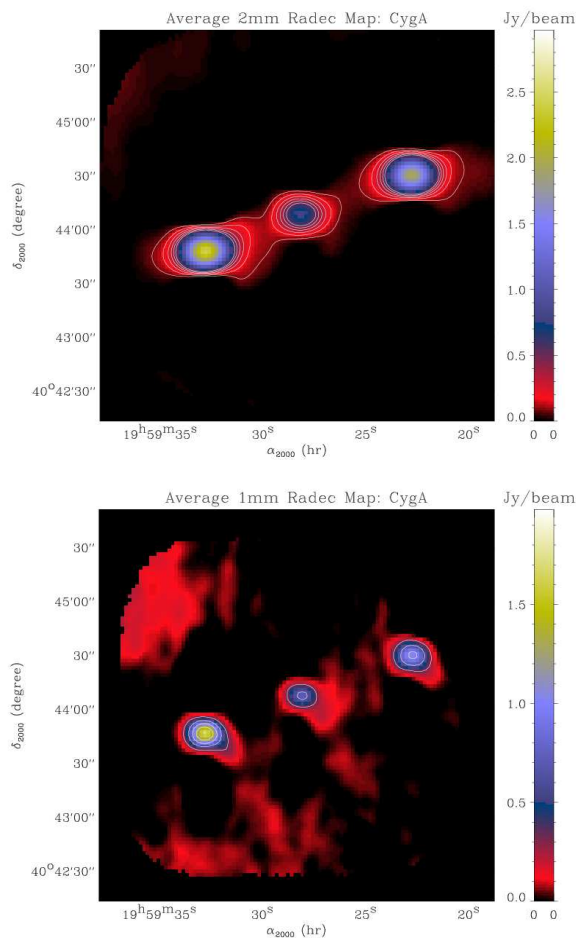}
    \caption{Maps at 2 and 1.4~mm of the radio source Cygnus A and its two radio lobes. All the
    three sources are detected at both wavelengths. Contours are at levels of
    $0.088\,\mathrm{Jy/beam}$ (resp. $0.332\,\mathrm{Jy/beam}$) and multiples at
    1.4 (resp. 2~mm). A Gaussian smoothing of 8 arcsecond has been applied to the
    maps.}
     \label{Fig10}
\end{figure}

In \figref{Fig10}, we show Cygnus~A (3C405), a well-known radio source with two
prominent radio lobes. The maps were obtained with two scans making a total
integration time of 2200~seconds. The central source flux is of
$0.76\pm0.09\,\mathrm{Jy}$ (resp. $0.89\pm0.04\,\mathrm{Jy}$) at 1.4~mm
(resp. 2~mm). The flux errors are dominated by photometric noise and not
detector noise.  The two radio lobes (named AB and DE) have a flux (taken at the
positions given by \cite{Wright2004}) of $1.63\pm0.09\,\mathrm{Jy}$ and
$1.65\pm0.08\,\mathrm{Jy}$ at 1.4~mm. At 2~mm, the measured fluxes are
$2.47\pm0.13\,\mathrm{Jy}$ and $2.36\pm0.10\,\mathrm{Jy}$. The 1.4~mm fluxes
can be compared with BIMA 1.3~mm interferometer observations by
\cite{Wright2004}. The BIMA fluxes of the nucleus, the AB and DE radio lobes
are 0.48, 0.54, and 0.97~Jy. The spectral dependance ($F_\nu \propto
\nu^{-1}$) as measured by \cite{Wright2004} is recovered on the two radio
lobes with the NIKA camera. Nevertheless, the NIKA fluxes are a factor 1.70
higher. We think that this an angular resolution effect. For example, using
the 1.1~mm flux measured by \cite{Eales1989} with a 19~arcsecond beam and
applying the above spectral dependence, we expect $0.74$, $1.1$ and
$1.5\,\mathrm{Jy}$ at 1.4~mm for the flux of the nucleus and the two radio
lobes which is in agreement with what we obtain.

\begin{figure}[h]
    \centering
    \includegraphics[width=.95\linewidth]{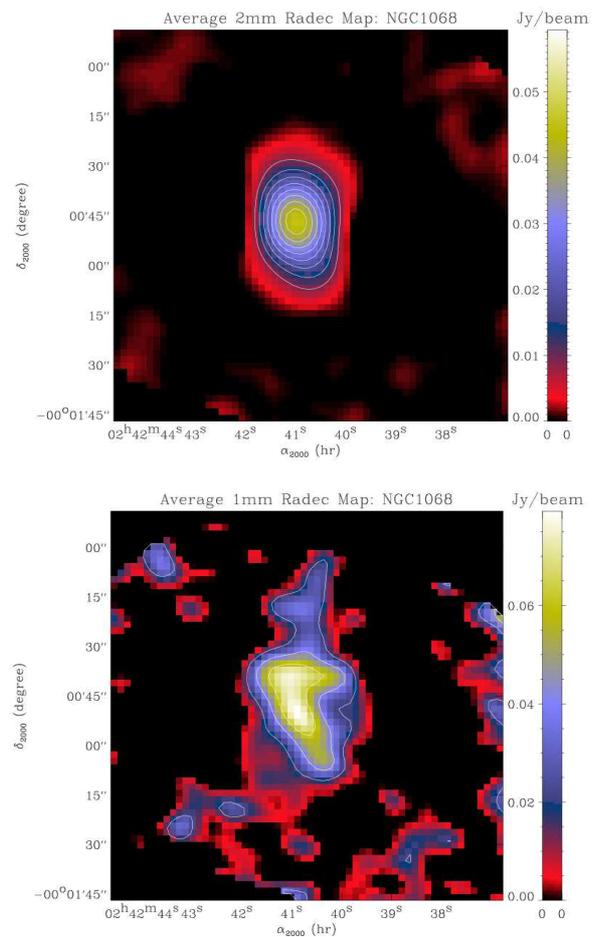}
    \caption{NIKA maps of NGC1068 galaxy at 2 and 1.4~mm wavelengths. The pixel size is 2
    arcseconds. The maps have been smoothed with an 8 arcsecond Gaussian. Contours
    are at the levels of $9.2 \,\mathrm{mJy/beam} \times (2, 3, ... 7)$ for the
    2~mm channel and 21, 42, $63 \,\mathrm{mJy/beam}$ for the 1.4~mm channel.}
     \label{Fig11}
\end{figure}

\figref{Fig11} shows a secure detection of NGC1068, a nearby galaxy with an
active galactic nucleus. The map is obtained with 5 scans and a total
integration time of 1260 seconds. Its flux is $142\pm25\,\mathrm{mJy}$ at 1.4~mm
and $66\pm3\,\mathrm{mJy}$ at 2~mm. This is the central flux measured with
a Gaussian of 12 and 19 arcseconds, respectively. For this map, a sky noise
decorrelation has been used, which is based on a linear regression with the
detector signals when off-source. The North East - South West extension at 1mm
is larger but aligned with interferometric IRAM Plateau de Bure
Interferometer (PdBI) observations (\cite{Krips2006}). The PdBI measured flux for
the core and the jet is $28\pm2\,\mathrm{mJy}$ at $231\,\mathrm{GHz}$. This
is smaller than the NIKA flux measurement which is more consistent with the
flux of $170\pm30\,\mathrm{mJy}$ measured by \cite{Thronson1987}, indicating
the presence of a diffuse extended component. This component is expected
(\cite{Hildebrand1977}), based on consideration of the far infrared spectral
energy distribution. It is likely to come from heated dust in the
circumnuclear region.

From deep integration on weak sources and using sky noise decorrelation, we
are able to derive the effective sensitivity of the camera, in the present
early state of data reduction and sky noise subtraction techniques.
We obtain a weak-source flux detectivity of 450 and $37,\mathrm{mJy}\cdot
s^{1/2}$ at 1.4 and 2~mm respectively. The 1.4~mm detectivity is
satisfactory for an initial 1.4~mm KID prototype. The 2~mm detectivity shows a
major improvement by a factor 3 with the best value obtained in the 2009 NIKA
run (\cite{Monfardini:29}). This detectivity is almost at the level of the
state-of-art APEX-SZ TES detectors (\cite{Schwan2010}) albeit obtained here
with a larger telescope. An NET of $6\mathrm{ mK} \cdot s^{1/2}$ is deduced
from this value. A sensitivity to the SZ effect can be obtained as $3 \cdot
10^{-5} hr^{1/2}$ in the $y$ Compton parameter for one beam, although this may
be complicated by the SZ extension. SZ measurements will be reported in a
later publication.
%END XAVIER CONTRIBUTION

%PART ON RAW S/N
\begin{figure}[h]
    \centering
    \includegraphics[width=.95\linewidth]{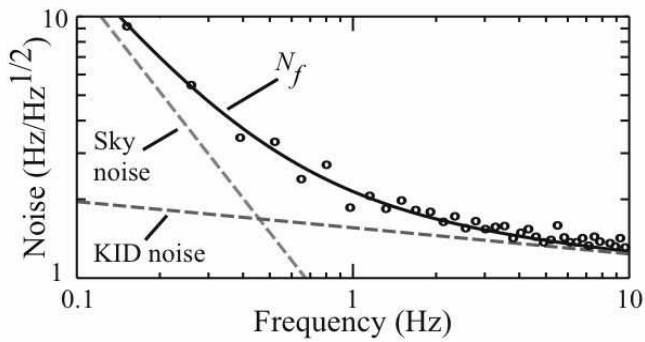}
    \caption{Frequency noise spectrum taken during a typical on-the-fly scan over the G34.3 galactic source.  A fit of the function in equation (4) yields the contribution to the noise due to low-frequency fluctuations from the source and sky ($\alpha=-1.35$) and the contribution at higher frequencies from the KIDs and the electronics ($\beta = -0.15$).  }
     \label{fig12}
\end{figure}

The individual detector performances were determined using noise spectra taken while on-telescope following the same strategy implemented with the sky simulator.  A typical noise spectrum, taken with the LEKID array during a standard on-the-fly scan, is shown in \figref{fig12}.  The data is fit using a sum of two power laws:
\begin{equation}
\label{4}
N_{f}(f) = A f^{\alpha} + B f^{\beta}
\end{equation}
The low-frequency fluctuations seen while observing are due to the expected source and sky noise ($\alpha=-1.35$) while the higher frequency fluctuations can be attributed to the detector noise ($\beta=-0.15$).  The detectors noise is relatively flat, in contrast with previous measurements where a $N_{\theta , f} \propto f^{-0.5}$ dependence was observed (\cite{Monfardini:29}, \cite{gao:102507}). This seems to indicate a substantial reduction, at least in the case of the LEKID array, of the intrinsic frequency noise due to random variations of the effective dielectric constant.  For the 220 GHz (1.4~mm) antenna-coupled KID array, the spectral slope does not contradict the cited model and exhibits the expected $f^{-0.5}$ dependence. We recognize however that dark measurements would be more suited to draw definitive conclusions concerning fundamental noise properties of the resonators.

Using the frequency response of Mars, a primary calibrator, and the noise spectra, we calculate a raw detector sensitivity of the order of 25 mJy/Hz$^{1/2}$ for the LEKID array and 180 mJy/Hz$^{1/2}$ for the antenna-coupled KID array. Both numbers refer to a single typical pixel and at 1 Hz. In terms of the optical NEP, with reasonable assumptions on the optical transmission chain and the pixel geometry, we estimate an average $2.3 \cdot 10^{-16}$ W/Hz$^{1/2}$ for the LEKID array and about $1.5 \cdot 10^{-15}$ W/Hz$^{1/2}$ for the antenna-coupled array, both at 1 Hz. These values include a 50\% factor which accounts for the impinging power reduction due to the polarizer. No further corrections have been applied to account for the detectors intrinsic optical efficiency. Millimeter-wave VNA measurements performed at room temperature on the LEKID array indicate a good radiation coupling, with a peak absorption (central frequency) exceeding 80\% (\cite{Roesch2010}). Three dimensional EM simulations suggest about 65\% for the antenna-coupled array optical absorption. The estimated NEP are, for both arrays, in good agreement with the preliminary values obtained using the sky simulator.
%END PART ON RAW S/N

%______________________________________________________________

\section{Conclusions and perspectives}

% History and overview of individual detector performance
Work on the NIKA instrument commenced in November 2008.  At that time, no mm-wave LEKID had been designed and fabricated while substantial research on mm-wave antenna-coupled KID designs had been conducted (\cite{Schlaerth2008}).  In only two years, NIKA has developed rapidly and the 150 GHz LEKID array sensitivity is now comparable to the existing IRAM instrument MAMBO2 at Pico Veleta.  While the sensitivity of the 220 GHz antenna-coupled array did not attain the target performance during this run, no significant barrier exists toward a substantial improvement in the near future.

For both arrays, further investigation is needed to achieve optimal photon-noise limited performance. Currently, a number of avenues are being pursued in order to achieve this goal:
\begin{enumerate}{}{}
\item New single-pixel geometries are being designed which promise to reduce pixel cross-talk, minimize detection volume, and absorb both polarizations.

\item Promising materials such as TiN are being characterized which may drastically improve the quantum efficiency of the detectors (\cite{Leduc:2010}).

\item The readout procedure continues to be optimized.  A poor photometric precision, of the order of 30\%, was obtained during this run.  Development of new models of the underlying superconducting physics is necessary to better understand the pixel response to incident radiation. A new off-line procedure is under development to include the $\partial\phi/\partial f$ non-linearity and thus improve the photometric accuracy. Also, leveraging experience acquired during previous bolometer array development, we are implementing a modulation-based readout procedure that should simplify and improve the photometry by providing the frequency response of every pixel in real-time.

%Nikel.... I like the name.
\item New readout electronics are being developed to further increase the multiplexing (MUX) factor of each readout cable.  The current instrument allows 112 pixels to be read out simultaneously in 233 MHz of bandwidth.  Compared with the first version of NIKA, this is a 6 fold improvement in the mux factor.  While a significant increase, digital electronics are readily available that can further improve the mux factor by more than an order of magnitude.  In order to harness this technology, we are developing a NIKA-dedicated readout system called NIKEL (NIKa ELectronics). A first generation NIKEL prototype, unused in the current measurement, achieved 128 channels over a bandwidth of 125 MHz (\cite{Bourrion2011}).  The next generation NIKEL system is under development and will achieve more than 256 channels over $\sim$400-500 MHz of bandwidth, using 12-bit analog-to-digital converters.

\item While greater magnetic shielding was utilized for the second generation NIKA system, magnetic field effects still limited the system performance.  In particular, the Earth's magnetic field was evident when changing the telescope azimuth angle.  Changes in elevations did not strongly affect the current detectors, but potentially limit the performance of higher sensitivity instruments.  Using a large current-carrying coil of wire, it is now possible to mimic the observed effects due to the Earth's field in the laboratory.  This new testing capability will allow a much more robust magnetic screening to be implemented and tested before installation at the telescope.

\item Improved baffling and filters are being designed to continue reducing spurious radiation.  Compared with the previous NIKA run, the current instrument reduced the stray light by more than a factor of two but stray light continues to degrade the detector performance.

\item NIKA will move to a cryogen free system.  While not affecting the system sensitivity, the use of a standard dilution cryostat is a practical limitation prohibiting long-term installation of NIKA at the Pico Veleta telescope.  The new cryogen-free cryostat has already been fabricated and is currently undergoing testing.

\end{enumerate}

The current version of NIKA demonstrates the potential for KIDs to operate in large, ground-based mm-wave instruments.  The NIKA project goal is to be a 6 arcminutes field-of-view, dual-band resident instrument at the 30-meter Pico Veleta telescope.  In order to preserve the telescope intrinsic angular resolution, a Nyquist sampling of the focal plane is targeted (0.50$\cdot$F$\lambda$).  This requires $\sim$1500 pixels at 150 GHz and $\sim$3000 pixels at 220 GHz.  As an intermediate step to achieving this specification, NIKA will be upgraded to a cryogen-free, dual-band instrument covering a field-of-view of about 3 arcminutes. Assuming the present $\approx$0.75$\cdot$F$\lambda$ sampling, a 224 pixel array working at 150 GHz and a 489 pixel array at 220 GHz are required. Assuming a MUX factor of 256, only three cold amplifiers and six coaxial cables are necessary.  A forthcoming measurement, implementing these system improvements is planned for the near future.

\begin{acknowledgements}
    We would like to thank Santiago Navarro, Juan Pe\~{n}alves, Fr\'{e}d\'{e}ric Damour, Carsten Kramer, David John, Juan Luis Santar\'{e}n, Denise Riquelme, Salvador Sanchez, Hans Ungerechts, Robert Zylka and all the IRAM staff for the excellent technical support during the run. We also acknowledge the technical staff at Institut N\'{e}el that have built the NIKA cryostat and participated in the electronics development, in particular Henri Rodenas, Gregory Garde, Anne Gerardin, Julien Minet and in general the Cryogenics and Electronics Groups.  This work was supported in part by grant ANR-09-JCJC-0021-01 of the French National Research Agency, the Nanosciences Foundation of Grenoble and R\'egion Rh\^one-Alpes (program CIBLE 2009). Part of the travel funds for the run have been provided by the French Minist\`{e}re des Affaires \'{e}trang\`{e}res et europ\'{e}ennes (PHC Alliance 2010). This work is supported, in the UK, by STFC. This research, and in particular A. Baryshev, was supported ERC starting Researcher Grant ERC-2009-StG 240602 TFPA. Akira Endo is financially supported by NWO (VENI grant 639.041.023) and the Netherlands Research School for Astronomy (NOVA). We acknowledge the crucial contributions of Ben Mazin (UCSB), Bruno Serfass (Berkeley) and the OSR (Open Source Readout for MKIDs) collaboration to the NIKA 2010 electronics.
\end{acknowledgements}

\end{document}